# SIMULATIONS OF THE FINITE TEMPERATURE QCD PHASE TRANSITION ON THE LATTICE [a]

KAZUYUKI KANAYA

*Center for Computational Physics and Institute of Physics,*
*University of Tsukuba, Tsukuba, Ibaraki 305, Japan*
*e-mail: kanaya@rccp.tsukuba.ac.jp*

I review the present status of nonperturbative studies on the nature of the finite temperature QCD transition on the lattice, with the special attention to the determination of the order of the transition.

## 1 Introduction

Unlike in the case of the usual formulation of renormalizable field theories in continuum space-time, where the UV cutoff is introduced through the perturbation theory, the lattice formulation [1] uses a finite lattice spacing, $a$, which acts as a UV cut-off independent of the validity of the perturbative expansions. This enables us to study nonperturbative properties of the theory, which are essential in the finite temperature physics of the quantum chromodynamics (QCD).

The continuum theory is defined in the limit of vanishing lattice spacing, $a = 0$, and infinite lattice volume. Before taking the continuum limit, the theory is finite and well defined, so that we can perform numerical simulations on super computers. Extrapolating the results to the continuum limit, we are discussing, for example, the hadron spectrum and decay constants to the accuracy of about 10%.

In this paper, I review the present status of numerical lattice calculations for the finite temperature QCD transition, with the special attention to the determination of the order of the transition taking the effects of dynamical quarks into account. For recent reviews on finite temperature QCD on the lattice, see Refs. 2 and 3.

---

[a] Review talk presented at the International Symposium on *Origin of Matter and Evolution of Galaxies in the Universe*, Jan. 18 – 20, 1996, Atami, Japan. To be published in the proceedings.

In nature, we know six flavors of quarks: u, d, s, c, b, and t. The lightest two quarks, u and d, are much lighter than the relevant energy scales whose typical magnitude can be given by the critical temperature $T_c$: $m_u$, $m_d \ll T_c$ (in the units $\hbar = c = k_B = 1$). On the other hand, the last three quarks, c, b, and t, are sufficiently heavy and are expected to play no appreciable roles to the thermal processes near $T_c$. In this review, I concentrate on the physically interesting cases of two and three flavors of degenerate light quarks ($N_F = 2$ and 3) and the more realistic case of two light and one heavy quarks ($N_F = 2 + 1$). The case $N_F = 2$ corresponds to the case that the third quark, the s quark, is much heavier than the relevant energy scale: $m_s \gg T_c$. The case $N_F = 3$ corresponds to the case $m_s \ll T_c$. In the following, we will see that the nature of the transition in the chiral limit (the limit of vanishing quark masses) for $N_F = 2$ is different from that for $N_F = 3$. Because $m_s \simeq 150$ – 200 MeV is just of the same order of magnitude as the expected values of $T_c$, this difference means that, in order to make a reliable prediction for the real world, we have to fine-tune the value of $m_s$ in the more realistic case $N_F = 2 + 1$. Unfortunately, the lattice simulations performed so far do not have that accuracy.

In the next section, I give a short introduction to the lattice formulation of QCD. In Sect. 3, I review recent studies on the QCD transition in the chiral limit. I then discuss the influence of the s quark in Sect. 4. A brief summary is given in Sect. 5.

## 2 QCD on the lattice

On a 4-dimensional Euclidian lattice, the gluon fields of QCD are provided by $3 \times 3$ complex matrices, $U_{x,\mu} \in \mathrm{SU}(3)$, living on the "links" connecting two neighboring sites, $x$ and $x + \hat{\mu}$, where $\hat{\mu}$ is the lattice unit vector in the $\mu$ direction. Near the continuum limit, $a \simeq 0$, the "link variable" $U_{x,\mu}$ is related to the conventional gluon field $A_\mu(x)$ by

$$U_{x,\mu} = \exp[igaA_\mu(x)], \qquad (1)$$

where $g$ is the gauge coupling constant. As in the continuum limit, quark fields on the lattice are given by Grassmann variables living on the sites.

The action on the lattice is chosen such that the continuum action is recovered when we let $a \to 0$ in the action. We also want to keep as much symmetries of the continuum action as possible on the lattice.

The minimal choice for the gluon action is the standard one-plaquette

action by Wilson,[4]

$$S_{gluon} = \beta \sum_{plaquettes} \frac{1}{3}\text{ReTr} U_{plaq} \qquad (2)$$

where $U_{plaq}$ is the ordered product of four $U_{x,\mu}$'s winding around a minimal square, the "plaquette", and $\beta = 6/g^2$. Due to the quantum fluctuations, the lattice spacing $a$ is a decreasing function of $\beta$ described by the renormalization group beta function. Therefore, the continuum limit corresponds to the limit $\beta = \infty$.

On the lattice, however, we have a freedom to introduce less local terms without affecting the continuum limit. By adjusting these additional coupling parameters, it is possible to accelerate the approach to the continuum limit. Such actions are called "improved actions". Importance of improvement is much stressed recently in many studies of lattice gauge theories. I will discuss an example for finite temperature physics later.

For the quark part, it is known that naive lattice discretizations of the continuum action have the "doubler problem": they necessarily contain additional fermionic degrees of freedom which survive in the continuum limit. It is inevitable to adopt a slightly complicated formulation on the lattice, abandoning some of simple features of the continuum theory.[5] Two conventional choices are the staggered and the Wilson fermions.

In the formulation of staggered quarks,[6] we introduce one component Grassmann field $\chi$ on each site. The 4 component Dirac field $\Psi$ is constructed collecting $\chi$ fields distributed on hypercube. Because a hypercube has $2^4$ sites, we end up with 4 flavors of Dirac fields. These 4 flavors are degenerate also in the continuum limit. Due to a flavor mixing interaction among these 4 flavors, the flavor/chiral symmetry $SU(N_F)_L \times SU(N_F)_R \times U(1)$ of massless QCD in the continuum limit[b] is explicitly broken down to $U(N_F/4)_L \times U(N_F/4)_R$ at $a > 0$. But, because at least a part of the chiral symmetry is preserved, the location of the chiral limit is protected by this symmetry in the coupling parameter space of the lattice action. This makes the choice of simulation parameters easier. A big flaw of this formulation is that $N_F$ must be a multiple of 4. For the most interesting cases $N_F = 2$ and 3, a usual trick is to modify the power of the fermionic determinant in the numerical path-integration by hand. This necessarily makes the action non-local, that sometimes poses conceptually difficult problems, as I discuss later. Practically, numerical results of hadron spectrum seem to suggest that the major lattice artifacts become quickly small at large $\beta$.

---

[b]The classical symmetry, $U(N_F)_L \times U(N_F)_R$, is broken down to this symmetry due to the quantum $U_A(1)$ anomaly.

Another conventional choice for the quark action is the Wilson fermion action.[7] In this formulation, we introduce four component Dirac field $\Psi$ on each site. Therefore the flavor symmetry is manifest also on the lattice. In order to avoid the doubler problem, however, we have to introduce the "Wilson term" to the action, which vanishes when we let $a \to 0$ naively in the action but explicitly breaks the chiral symmetry at $a > 0$. The numerical studies show that the effects of this chiral violation are small and are well under control.

In the continuum limit, we expect that both Wilson and staggered quarks give the same results, at least for $N_F = 4\times$ integer. Off the continuum limit, they are in general different, and the difference provides us with a measure of the quality of predictions from the lattice.

## 3  Finite temperature chiral transition in QCD

At low temperatures, QCD has three characteristic properties: confinement of quarks and gluons, spontaneous breakdown of the chiral symmetry, and the asymptotic freedom. The first two are IR properties and may be modified by temperature. Actually, the study of lattice QCD has shown that there exists a finite temperature transition between the low temperature hadronic phase where quarks are confined and the chiral symmetry is spontaneously broken, and the high temperature quark gluon plasma (QGP) phase where quarks are deconfined and the chiral symmetry is restored. Lattice simulations further suggest that the confinement-deconfinement transition and the recovery of the chiral transition take place at the same temperature.

In QCD we have several adjustable parameters such as quark masses and the number of flavors. Phenomenological interests are, of course, concentrated on the nature of the transition in a particular case where the quark masses have their physical values. This, however, requires a rather delicate fine-tuning of parameters in a lattice simulation. Therefore, it is important to make clear the nature of the transition in limiting cases.

One of these interesting limiting cases of QCD is the limit of the SU(3) pure gauge theory, that corresponds to the case of infinitely heavy quarks, $m_q \gg T_c$. In this limit, QCD has a global Z(3) symmetry, which is spontaneously broken when quarks are deconfined. This suggests that the deconfining transition of the SU(3) gauge theory belongs to the same universality class as the order-disorder transition of three dimensional Z(3) spin models such as the 3-state Potts model. Because the transition is of first order in these spin models, we expect that the deconfinement transition is also of first order. Precise lattice studies applying the finite size scaling analysis confirmed this expectation.[8,9]

The success of the universality argument in the pure gauge limit encour-ages us to apply a similar idea to more realistic cases with dynamical quarks. Here, the Z(3) symmetry of the pure gauge theory is explicitly broken by the quark mass term. Therefore, we cannot use this symmetry for universality argument. In the massless limit (chiral limit), however, we have the spontaneously broken chiral symmetry that is expected to recover at $T > T_c$. Based on a study of an effective $\sigma$ model which respects the chiral symmetry of QCD, Pisarski and Wilczek[10] discussed that the transition in the chiral limit (chiral transition) of QCD is of first order for $N_F \geq 3$. This was confirmed by numerical simulations both with staggered[11] and with Wilson quarks.[12] (See Ref. 2 for a more complete list of references.)

Table 1: Critical exponents of the three dimensional O(4)[14] and O(2)[15] Heisenberg models, and $N_F = 2$ QCD with staggered quarks.[16]

|  | O(4) | O(2) | Karsch-Laermann |  |
|---|---|---|---|---|
| $1/\beta\delta$ | 0.537(7) | 0.602(2) | 0.77(14) | $\beta_c$ |
| $1/\delta$ | 0.2061(9) | 0.2072(3) | $0.21 - 0.26$ | chiral cumulant |
| $1 - 1/\delta$ | 0.7939(9) | 0.7928(3) | 0.79(4) | $\chi_m$ |
| $(1-\beta)/\beta\delta$ | 0.331(7) | 0.394(2) | 0.65(7) | $\chi_t$ |
| $\alpha/\beta\delta$ | $-0.13(3)$ | $-0.003(4)$ | $-0.07 - +0.34$ | specific heat |

For two flavors, theoretical studies do not have a definite prediction about the order of the chiral transition: the transition is either of second order or of first order depending on the strength of the $U_A(1)$ anomaly term at the transition temperature.[10] In case that the chiral transition is of second order, the universality argument suggests that it will belong to the same universality class as the order-disorder transition of the three dimensional O(4) Heisenberg model,[13] which shows well-studied critical behaviors. This provides us with several useful scaling properties among physical quantities around the chiral transition that allow us to make clear the order of the transition by numerical simulations.

Scaling properties are described in terms of critical exponents. Consider a spin model at temperature $T$ near the transition temperature $T_c$ and at small external magnetic field $h$. The exponents $\alpha$, $\beta$, $\gamma$, and $\delta$ are defined for the specific heat $C$, the magnetization $M$, and the magnetic susceptibility $\chi$ by

$$C(t, h=0) \sim |t|^{-\alpha} \qquad (3)$$
$$M(t, h=0) \sim |t|^{\beta} \quad (t<0) \qquad (4)$$
$$\chi(t, h=0) \sim t^{-\gamma} \quad (t>0) \qquad (5)$$
$$M(t=0, h) \sim h^{1/\delta} \qquad (6)$$

where $t = [T - T_c(h=0)]/T_c(h=0)$ is the reduced temperature. These critical exponents satisfy the following two scaling relations: $\alpha = 2 - \beta(\delta + 1)$ and $\gamma = \beta(\delta - 1)$, so that only two of the four exponents are independent. We also expect that magnetization $M$ near the transition point can be described by a single scaling function:

$$M = h^{1/\delta} f(t/h^{1/\beta\delta}). \tag{7}$$

Some O(4) exponents are listed in Table 1.

In QCD, $h$ corresponds to the quark mass and $M$ corresponds to the chiral condensate (the chiral order parameter) $\langle \bar{\Psi}\Psi \rangle$. On a lattice with the linear lattice size $N_t$ in the time direction, temperature is given by $T = 1/[N_t a(\beta)]$, where the lattice spacing $a$ is a decreasing function of $\beta$, as discussed before. In simulations with $N_t$ fied, larger $\beta$ corresponds to larger $T$. The continuum limit at a given value of $T$ is achieved by letting $N_t \to \infty$ and $a \to 0$, simultaneously.

The conjecture of the O(4) scaling has been tested first for staggered quarks by F. Karsch and E. Laermann on an $N_t = 4$ lattice.[16] The last two columns in Table 1 show their numerical results for critical exponents. The result for $\delta$ is consistent with the O(4) value, while the results for $\alpha$ and $\beta$ are in disagreement with the O(4) exponents. It was argued that the discrepancies of $\alpha$ and $\beta$ might be caused by the small lattice size simulated. C. DeTar tested the scaling relation (7) using data for $\langle \bar{\Psi}\Psi \rangle$ from two flavor staggered quarks [17] and found that data are roughly consistent with the O(4) saling (and also with the O(2) scaling, discussed below). With staggered quarks, however, several caveats are in order because $N_F = 2$ staggered quarks are realized on the lattice by introducing a fractional exponent to the fermionic determinant of the $N_F = 4$ staggered quarks. Therefore, (i) the symmetry in the chiral limit on a lattice with finite lattice spacing is not the O(4) but O(2) of $N_F = 4$ staggered quarks, and (ii) the action is not local. The correct continuum chiral limit with the O(4) symmetry will be obtained only when we first take the continuum limit $\beta \to \infty$ and then take the chiral limit. Choosing a too small $m_q$ compared with the lattice spacing may lead us either to wrong O(2) exponents, or to some non-universal behavior due to the lack of locality. In order to conclude the O(4) scaling among these possibilities, we clearly need data with better accuracy on larger lattices. At least, the appearance of non-trivial exponents strongly suggests that the transition is of second order, although the lattice artifact may be large as we cannot see the clear O(4) scaling.

In the case of Wilson quarks, the lattice action lacks chiral symmetry due to the Wilson term. This introduces possible $O(a)$ effects to physical quantities and relations characteristic for chiral symmetry. However, when we are close enough to the continuum limit, we expect that the symmetry breaking effects become sufficiently weak so that we see the O(4) scaling when the chiral transition is of second order.[2] With the standard action, we found that the effects of the Wilson term are not negligible on lattices available with the present power of computers.[3] We therefore applied an improved action to study this issue.[18] We first confirmed that the lattice artifact from the Wilson term is small with our improved action. We then studied the scaling function in (7) with the identification $h = 2m_q a$ and $t = \beta - \beta_{ct}$, where $\beta_{ct}$ is the chiral transition point. Fixing the exponents to O(4) values given in Table 1, we adjust $\beta_{ct}$ to obtain the best fit. Our result is shown in Fig. 1. We find that the scaling ansatz works remarkably well. This result is quite nontrivial because a slight shift of the exponents to other values suggested, for example, from the meanfield theory makes the fit apparently worse.

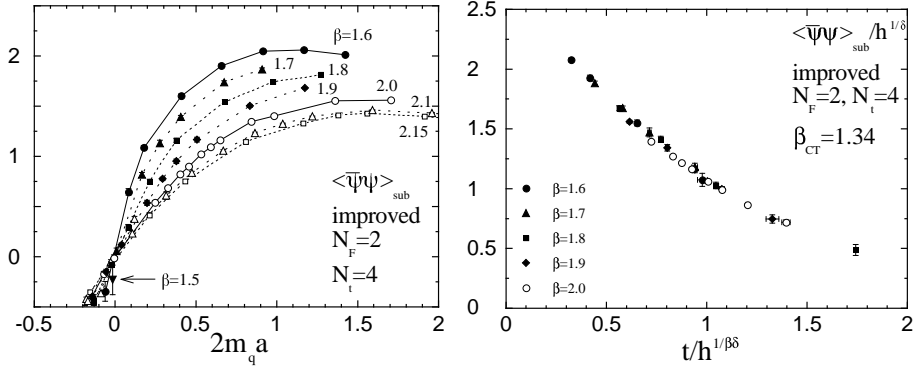

Figure 1: (a) Chiral condensate as a function of $2m_q a$ for Wilson quarks with a RG improved action on an $8^3 \times 4$ lattice.[18] (b) Best fit for the scaling function with O(4) exponents. The plot contains all data of Fig. 1(a) within the range $0 < 2m_q a < 0.9$ and $\beta \leq 2.0$.

## 4 Influence of the strange quark

In the previous section, we have seen that the finite temperature transition is of second order in the limit of massless quarks in $N_F = 2$ QCD, while it is of first order for $N_F \geq 3$. Off the chiral limit, the first order transition smoothens into a crossover at sufficiently large $m_q$. Therefore, the nature of the transition sensitively depends on $N_F$ and $m_q$. This means that, in order to study the nature of the transition in the real world, we should include the s quark properly whose mass $m_s$ is of the same order of magnitude as the transition temperature $T_c \simeq 100 - 200$ MeV.

Following Brown et al.,[11] we summarize in Fig. 2 what we expect about the nature of the finite temperature transition as a function of quark masses, neglecting the mass difference among u and d quarks ($N_F = 2+1$). When all quarks are heavy, the transition is of first order as observed in the SU(3) pure gauge theory. The limit $m_s = \infty$ corresponds to the case $N_F = 2$ discussed in Sect. 3 where we found strong evidence for second order transition at $m_{ud} = 0$. When the chiral transition is of second order, we expect that it turns into an analytic crossover at nonvanishing quark masses. For $m_{ud} = m_s$ ($N_F = 3$), the transition is of first order in the chiral limit. Therefore, on the axis $m_{ud} = 0$, we have a tricritical point $m_s^*$ where the second order transition at large $m_s$ turns into first order [13]. For $m_s > m_s^*$, the second order transition line follows the $m_{ud} = 0$ axis and, for $m_s < m_s^*$, is suggested [19] to deviate from the vertical axis according to $m_{ud} \propto (m_s^* - m_s)^{5/2}$.

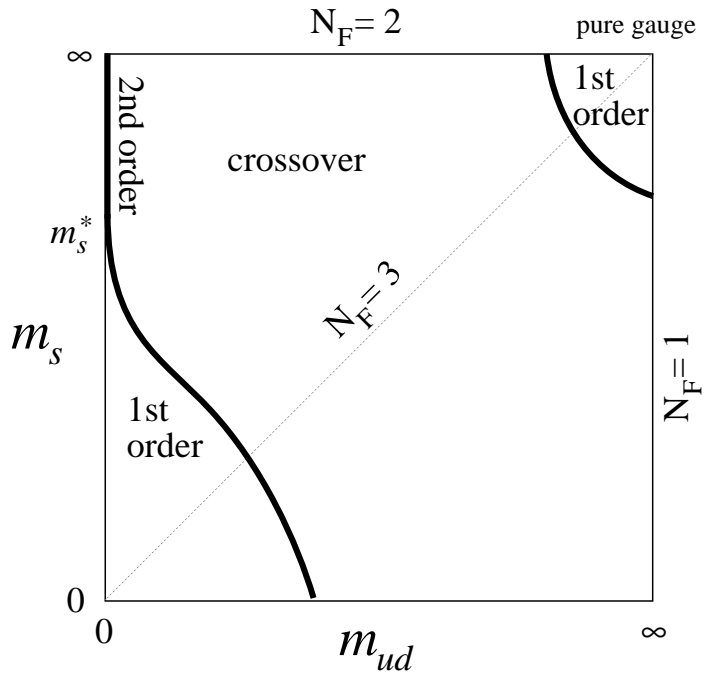

Figure 2: Map of expected nature of the QCD transition for $N_F = 2+1$ QCD as a function of the u and d quark mass $m_{ud}$ and the s quark mass $m_s$.

Our main goal is to determine the position of the physical point in this map.

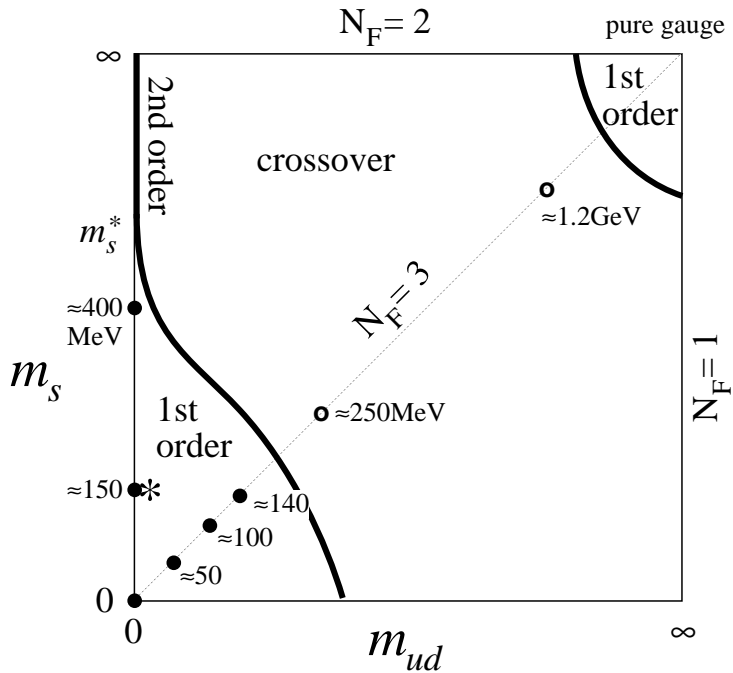

Figure 3: Same as Fig. 2 with the results of simulations with Wilson quarks using the standard action.[20] First order signals are observed at the points marked by filled circles, while no clear two state signals are found at the points represented by the open circles. The values of quark mass in physical units are computed using $a^{-1} \sim 0.8$ GeV for $\beta \leq 4.7$ and $a^{-1} \sim 1.0(1.8)$ GeV for $\beta = 5.0(5.5)$ determined by $m_\rho(T=0) = 770$ MeV. The real world corresponds to the point marked by the star.

Using staggered quarks, the Columbia group studied this issue five years ago on $N_t = 4$ lattices.[11] For the degenerate $N_F = 3$ case, $m_{ud} = m_s \equiv m_q$, they found a first order signal for $m_q a = 0.025$ at $\beta = 5.132$. For $N_F = 2 + 1$, they obtained a time history suggesting a crossover for $m_{ud} a = 0.025$, $m_s a = 0.1$ at $\beta = 5.171$. Their study of hadron spectrum at this point gives the value of $m_K/m_\rho$ smaller than the experimental value, suggesting that their $m_s$ is smaller than its physical value. At the same time, their large $m_\pi/m_\rho$ suggests that their $m_{ud}$ is larger than the physical value. This implies that the physical point is located in the crossover region unless the second order transition line, which has a sharp $m_{ud}$ dependence near $m_s^*$ (cf. Fig. 2), crosses between the physical point and the simulation point.

Recently, we studied this issue with Wilson quarks using the standard action on $N_t = 4$ lattices.[20] For $N_F = 3$, we performed simulations increasing the quark mass at the transition temperature and found that the first order signal observed in the chiral limit [12] persists for the cases corresponding to $m_q \lesssim 140$ MeV, while no clear two state signals are observed for $m_q \gtrsim 250$ MeV, where the physical s quark mass, giving $m_\phi = 1.02$ GeV, is about 150 MeV with our normalization of $m_q$. For $N_F = 2 + 1$, we observed first order signals for $m_{ud} \sim 0$ at both $m_s \sim 150$ and 400 MeV. Our study of hadron spectroscopy for $N_F = 2 + 1$ shows that $m_\phi \sim 1.03(5)$ GeV at the simulation point with $m_s \sim 150$ MeV, verifying that this simulation point is very close to the physical point in this sense. Our results on the nature of the transition are summarized in Fig. 3. The physical point clearly falls in the first order region.

Although both staggered and Wilson simulations give a phase structure qualitatively consistent with Fig. 2, Wilson quarks tend to give larger values for critical quark masses (measured by $m_\phi/m_\rho$ etc.) than those with staggered quarks. This leads to the difference in the conclusions about the location of the physical point. in Fig. 2. On the other hand, both of these studies discuss that the deviation from the continuum limit is large at $N_t = 4$ where these simulations are done: sizable deviations from the experimental values are observed in several physical quantities. We should certainly make a calculation at larger $N_t$ or with an improved action in order to draw a definite conclusion about the nature of the QCD transition in the real world. At present, increasing $N_t$ is quite painful. Therefore, we began to study these issues applying the improved action which was successful for $N_F = 2$. Some preliminary results were reported,[18] so far with no conclusion about the values for critical quark masses.

## 5 Conclusions

Results from lattice QCD mean that they are derived from the first principles of the theory of strong interaction. They, however, require careful extrapolations to the continuum limit. Numerical simulations of lattice QCD show strong evidences that the QCD transition in the massless quark limit is of first order for $N_F \geq 3$, while, for $N_F = 2$, it is of second order universal to the order-disorder transition of the three dimensional O(4) Heisenberg model. These results are in accordance with the theoretical expectations based on the universality argument. In order to make a prediction to the real world, we have to fine-tune the s quark mass to its physical value. Unfortunately, the present lattice simulations do not have that accuracy yet. We hope to solve this problem by new simulations applying improved actions, so that we may soon be able to make more definite predictions about the natute of the finite temperature QCD transition at the early Universe.

**Acknowledgments**

I am grateful to my collaborators, Y. Iwasaki, S. Kaya, S. Sakai, and T. Yoshié for their support. This work is in part supported by the Grant-in-Aid of Ministry of Education, Science and Culture (Nos.07NP0401 and 07640376).

**References**

1. For introduction of lattice gauge theories, see M. Creutz, *Quarks, Gluons and Lattices* (Cambridge University Press, Cambridge, 1988); I. Montvay and G. Münster, *Field Theory on the Lattice* (Cambridge University Press, Cambridge, 1993).
2. K. Kanaya, in *Lattice '95*, Proceedings of the International Symposium on Lattice Field Theory [Nucl. Phys. B (Proc. Suppl.) **47**, to be published (1996)].
3. C. DeTar, in *Lattice '94* [Nucl. Phys. B (Proc. Suppl.) **42**, 73 (1995)]; Y. Iwasaki, *ibid.*, 96 (1995).
4. K.G. Wilson, Phys. Rev. **D10**, 2445 (1974).
5. H.B. Nielsen and M. Ninomiya, Nucl. Phys. **B185**, 20 (1981); *ibid.* **B193**, 173 (1981).
6. J. Kogut and L. Susskind, Phys. Rev. **D11**, 395 (1975); L. Susskind, *ibid.* **D16**, 3031 (1977).
7. K.G. Wilson, in *New Phenomena in Subnuclear Physics*, ed. A. Zichichi (Plenum, New York, 1977).
8. For review, see A. Ukawa, Nucl. Phys. B (Proc. Suppl) **17**, 118 (1990).
9. Y. Iwasaki *et al.*, Phys. Rev. **D46**, 4657 (1992).
10. R. Pisarski and F. Wilczek, Phys. Rev. **D29**, 338 (1984);
11. F.R. Brown, *et al.*, Phys. Rev. Lett. 65(1990) 2491.
12. Y. Iwasaki, K. Kanaya, S. Sakai and T. Yoshié, Z. Phys. **C**, to be published (1996); Nucl. Phys. B (Proc. Suppl.) **30**, 327 (1993); *ibid.* **34**, 314 (1994).
13. F. Wilczek, Int. J. Mod. Phys. **A7**, 3991 (1992); K. Rajagopal and F. Wilczek, Nucl. Phys. **B399**, 395 (1993).
14. K. Kanaya and S. Kaya, Phys. Rev. **D51**, 2404 (1995).
15. J.C. Le Guillou and J. Zinn-Justin, Phys. Rev. **B21**, 3976 (1980); J. Phys. Lett. (Paris) **46**, L-137 (1985).

16. F. Karsch, Phys. Rev. **D49**, 3791 (1994); F. Karsch and E. Laermann, *ibid.* **D50**, 6954 (1994).
17. C. DeTar in Ref. 3
18. Y. Iwasaki, K. Kanaya, S. Sakai and T. Yoshié, Nucl. Phys. **B (Proc. Suppl.) 42**, 502 (1995); Y. Iwasaki, K. Kanaya, S. Kaya, S. Sakai and T. Yoshié, *ibid.* **47**, to be published (1996).
19. K. Rajagopal, in *Quark-Gluon Plasma 2*, ed. R. Hwa, World Scientific, 1995.
20. Y. Iwasaki, K. Kanaya, S. Kaya, S. Sakai and T. Yoshié, Z. Phys. **C**, to be published (1996); Nucl. Phys. B (Proc. Suppl.) **42**, 499 (1995).